\documentclass[preprint,12pt]{elsarticle}
\usepackage{amssymb}

\usepackage{amsthm}
\theoremstyle{definition}
\newtheorem{definition}{Definition}[section]
\newtheorem{corollary}{Corollary}[section]
\newtheorem{remark}{Remark}[section]
\theoremstyle{plain}
\newtheorem{theorem}{Theorem}[section]

\journal{Journal of Functional Analysis}

\begin{document}

\begin{frontmatter}

\title{Quasi-Feynman formulas -- a method of obtaining the evolution operator for the Schr\"{o}dinger equation}

\author{Ivan\,D.\,Remizov}

\address{Bauman Moscow State Technical University\\
Rubtsovskaya nab., 2/18, office 1027, 105005 Moscow, Russia\\
Lobachevsky Nizhny Novgorod State University\\
Prospekt Gagarina, 23, Nizhny Novgorod, 603950, Russia}

\ead{ivan.remizov@gmail.com}

\begin{abstract}

For a densely defined self-adjoint operator $\mathcal{H}$ in Hilbert space $\mathcal{F}$ the operator $\exp(-it\mathcal{H})$ is the evolution operator for the Schr\"odinger equation $i\psi'_t=\mathcal{H}\psi$, i.e. if $\psi(0,x)=\psi_0(x)$ then $\psi(t,x)=(\exp(-it\mathcal{H})\psi_0)(x)$ for $x\in Q.$ The space $\mathcal{F}$ here is the space of wave functions $\psi$ defined on an abstract space $Q$, the configuration space of a quantum system, and $\mathcal{H}$ is the Hamiltonian of the system. 
In this paper the operator $\exp(-it\mathcal{H})$ for all real values of $t$ is expressed in terms of the family of self-adjoint bounded operators $S(t), t\geq 0$, which is Chernoff-tangent to the operator $-\mathcal{H}$. One can take $S(t)=\exp(-t\mathcal{H})$, or use other, simple families $S$ that are listed in the paper. The main theorem is proven on the level of semigroups of bounded operators in $\mathcal{F}$ so it can be used in a wider context due to its generality. Two examples of application are provided.

\end{abstract}


\end{frontmatter}

\newpage
\normalsize            

\footnotesize
\tableofcontents             
\normalsize            
						
\section{Introduction}

A Feynman formula (in the sense of Smolyanov \cite{STT}) is a representation of a function  as the limit of a multiple integral where the multiplicity tends to infinity. Usually this function is a solution to the Cauchy problem for a partial differential equation (PDE).  In this paper we introduce a more general concept: 

\begin{definition}\label{qFFdef} A quasi-Feynman formula is a representation of a function in a form which includes multiple integrals of an infinitely increasing multiplicity.\end{definition} 

The difference with a Feynman formula is that in a quasi-Feynman formula summation and other functions/operations may be used while in a Feynman formula only the limit of a multiple integral where the multiplicity tends to infinity is allowed.  Both Feynman formulas and quasi-Feynman formulas approximate Feynman path integrals.  

Formula (\ref{FFeyn2}) and other formulas from theorem \ref{trick} are examples 
of quasi-Feynman formulas for the case when the (later discussed) family $(S(t))_{t\geq 0}$ consists of integral operators; the obtained formulas give the exact solution to the Cauchy problem for the Schr\"{o}dinger equation. 

It is known that the solution to the Cauchy problem for the Schr\"odinger equation  $i\psi'_t(t,x)=\mathcal{H}\psi(t,x), \psi(0,x)=\psi_0(x)$ is given by the formula
$\psi(t,x)= (\exp(-it\mathcal{H})\psi_0)(x)$; the evolution operator 
$\exp(-it\mathcal{H})$ is a one-dimensional (parametrized by $t\in\mathbb{R}$) group of unitary operators in Hilbert space. Because of the quantum mechanical significance,
the properties of $\exp(-it\mathcal{H})$ have been extensively studied. Research topics include e.g. exact solutions to the Cauchy problem, asymptotic behavior, estimates, related spatio-temporal structures, wave traveling, boundary conditions and other. Some of the recent papers related to the Cauchy problem solution study are \cite{HW, Ord, Nak, WZ, CorN, CNR,IN, Maz,Ah}.

In this paper we propose a method of obtaining formulas that express 
$\exp(-it\mathcal{H})$ in terms of the coefficients of the operator $\mathcal{H}.$ The solution is obtained in the form of a quasi-Feynman formula. Quasi-Feynman formulas are easier to obtain (compared with Feynman formulas) but they provide lengthier approximation expressions.

Suppose that function $u(t,x)$ is the solution for the following Cauchy problem: $u'_t=Lu, u(0,x)=u_0(x)$. The expression $$u(t,x)=\lim_{n\to\infty}\underbrace{\int_E\dots\int_E}_{n}\dots dx_1\dots dx_n$$
is called a Lagrangian Feynman formula if $E$ is a configuration space for the dynamical system that is described by the equation $u'_t=Lu$; it is called a Hamiltonian Feynman formula if $E$ is a phase space for the same system. For the first time Lagrangian Feynman formulas appeared in the paper by R.\,P.~Feynman \cite{F1} in 1948, who postulated them without proof. The proof based on the Trotter product formula was provided by E.~Nelson \cite{Nel} in 1964. Hamiltonian Feynman formulas were presented in Feynman's paper \cite{F2} in 1951, but the proof (based on the Chernoff theorem) was published only in 2002 by O.\,G.~Smolyanov, A.\,G.~Tokarev and A.~Truman \cite{STT}.

Note that the terminology can be obscure: we have Feynman integral, Feynman (pseudo)measure, Feynman formulas, Feynman-Kac formulas -- all these are different objects, and different authors define them in different ways.

One of the ways of obtaining and proving Feynman formulas is to use a one-parameter, strongly continuous semigroup of bounded linear operators (i.e. a $C_0$-semigroup, definition \ref{C0sem} below) as a solution-providing object, and the Chernoff theorem (theorem \ref{FormulaChernova} below) as the main technical tool to deal with the $C_0$-semigroup. The Chernoff theorem states that to obtain an explicit formula for a $C_0$-semigroup, it is enough to find a one-parameter family of bounded linear operators that is Chernoff-equivalent (definition \ref{Cheq} below) to the $C_0$-semigroup. So the task of solving the Cauchy problem for an evolutionary PDE is reformulated as the task of finding an appropriate family of operators. In all known examples, families of integral operators are used, and the Chernoff theorem requires to compose them many times, this is how multiple integrals in Feynman formulas arise in this approach. The history of research in this particular direction and a sketch of results obtained up to 2009 one can find in \cite{SmHist}; see also the overview \cite{SmSchrHist} dedicated to Feynman formulas for a Schr\"{o}dinger semigroup (2011). The most recent (but not complete) overview is \cite{Butko4} (2014, in Russian). The advances achieved employing this idea can be found in papers by Ya.\,A.~Butko (now Kinderknecht), M.\,S.~Buzinov, V.\,A.~Dubravina, A.\,V.~Du\-ryagin, A.\,S.~Plya\-shechnik, V.\,Zh.~Sakbaev, N.\,N.~Shamarov, O.\,G.~Smolyanov and in references therein. Some of the relevant papers are \cite{Remizov1, BGS2010, Plya2, BB, Bmnog, Plya1, Butko1, Butko2, OSS}. 

It is known that constructing
such Chernoff-equivalent families for a Schr\"odinger equation is much more
difficult than doing the same for a heat equation. O.\,G.~Smolyanov and his group  have constructed Chernoff-equivalent families
that provide the solution to the heat equation in many cases. Now this material can be used to solve the Cauchy problem for the Schr\"odinger quation.

In this paper we propose a specific family of operators that is Chernoff-equivalent to the evolution operator family 
$(\exp(-it\mathcal{H}))_{t\in\mathbb{R}}$ for the Schr\"odinger equation.  The family reads as 
$R(t)=\exp[i(S(t)-I)]$ and is a source of the quasi-Feynman formulas
that provide $\exp(-it\mathcal{H})$. The operator $R(t)$ that is introduced depends on the operator $\mathcal{H}$ via the operator $S(t)$; namely, we ask the family $(S(t))_{t\geq 0}$ to be Chernoff-tangent (definition \ref{weakCher} below) to the operator $-\mathcal{H}$. Such families $(S(t))_{t\geq 0}$ are known for a wide range of operators $\mathcal{H}$. The main theorem of the paper (theorem \ref{trick}) can be viewed as an analogue or a generalization of the Trotter product formula and the Chernoff product formula for the Schr\"odinger equation.

With the method presented the difficulty of solving the Cauchy problem for a Schr\"{o}dinger equation reduces twice: we need to construct a less difficult (Chernoff-tangent) family for a less difficult (heat) equation. This technique deals with the semigroups and operator families only, so it works for a large class of Hamiltonians describing dynamics in a large class of configuration spaces (two examples are shown in the article).

The method presented opens several challenging questions -- for example, it possibly may provide better approximations than Feynman formulas do, but this requires a further study, see remarks \ref{skorostskhod} and \ref{turkom} above.

\section{Preliminaries}In this section the essential background in $C_0$-(semi)group theory is provided.

\begin{definition}\label{C0sem}Let $\mathcal{F}$ be a Banach space over the field $\mathbb{C}$. Let $\mathcal{L}(\mathcal{F})$ be a set of all bounded linear operators in $\mathcal{F}$. Suppose we have a mapping $V\colon [0,+\infty)\to \mathcal{L}(\mathcal{F}),$ i.e. $V(t)$ is a bounded linear operator $V(t)\colon \mathcal{F}\to \mathcal{F}$ for each $t\geq 0.$ The mapping $V$ is called a \textit{$C_0$-semigroup}, or \textit{a strongly continuous one-parameter semigroup} if it satisfies the following conditions: 

1) $V(0)$ is the identity operator $I$, i.e. $\forall \varphi\in \mathcal{F}: V(0)\varphi=\varphi;$ 

2) $V$ maps the addition of numbers in $[0,+\infty)$ into the composition of operators in $\mathcal{L}(\mathcal{F})$, i.e. $\forall t\geq 0,\forall s\geq 0: V(t+s)=V(t)\circ V(s),$ where for each $\varphi\in\mathcal{F}$ the notation $(A\circ B)(\varphi)=A(B(\varphi))$ is used;

3) $V$ is continuous with respect to the strong operator topology in $\mathcal{L}(\mathcal{F})$, i.e. $\forall \varphi\in \mathcal{F}$ function $t\longmapsto V(t)\varphi$ is continuous as a mapping $[0,+\infty)\to \mathcal{F}.$

The definition of a \textit{$C_0$-group} is obtained by the substitution of $[0,+\infty)$ by $\mathbb{R}$ in the paragraph above.
\end{definition}

If $(V(t))_{t\geq 0}$ is a $C_0$-semigroup in Banach space $\mathcal{F}$, then the set $$\left\{\varphi\in \mathcal{F}: \exists \lim_{t\to +0}\frac{V(t)\varphi-\varphi}{t}\right\}\stackrel{denote}{=}Dom(L)$$ is dense in $\mathcal{F}$. The operator $L$ defined on the domain $Dom(L)$ by the equality $$L\varphi=\lim_{t\to +0}\frac{V(t)\varphi-\varphi}{t}$$ is called an infinitesimal generator (or just generator to make it shorter) of the $C_0$-semigroup $(V(t))_{t\geq 0}$. The generator is a closed linear operator that defines the $C_0$-semigroup uniquely, and the notation $V(t)=e^{tL}$ is used. If $L$ is a bounded operator and $Dom(L)=\mathcal{F}$ then $e^{tL}$ is indeed the exponent defined by the power series $e^{tL}=\sum_{k=0}^\infty\frac{t^kL^k}{k!}$ converging with respect to the norm topology in $\mathcal{L}(\mathcal{F})$. In most interesting cases the generator is an unbounded differential operator such as Laplacian $\Delta$.

One of the reasons for the study of $C_0$-semigroups is their connection with differential equations. If $Q$ is a set, then the function $u\colon [0,+\infty)\times Q\to \mathbb{C}$, $u\colon (t,x)\longmapsto u(t,x)$ of two variables $(t,x)$ can be considered as a function $u\colon t\longmapsto [x\longmapsto u(t,x)]$ of one variable $t$
with values in the space of functions of the variable $x$. If $u(t,\cdot)\in\mathcal{F}$ then one can define $Lu(t,x)=(Lu(t,\cdot))(x).$ If there exists a $C_0$-semigroup $(e^{tL})_{t\geq 0}$ then the Cauchy problem 
$$
\left\{ \begin{array}{ll}
 u'_t(t,x)=Lu(t,x) \ \mathrm{ for }\ t>0, x\in Q\\
 u(0,x)=u_0(x)\ \mathrm{ for } \ x\in Q
  \end{array} \right.
$$
has a unique (in sense of $\mathcal{F}$, where $u(t,\cdot)\in\mathcal{F}$ for every $t\geq 0$) solution $u(t,x)=(e^{tL}u_0)(x)$ depending on $u_0$ continuously. See  \cite{Pazy, EN1} for details. Note that if there exists a strongly continuous group $(e^{tL})_{t\in\mathbb{R}}$ then in the Cauchy problem the equation $u'_t(t,x)=Lu(t,x)$ can be considered not only for $t>0$, but for $t\in\mathbb{R}$, and the solution is provided by the same formula $u(t,x)=(e^{tL}u_0)(x)$.

The following theorem implies the existence and uniqueness of the solution for the Cauchy problem for the Schr\"{o}dinger equation.

\begin{theorem} (\textsc{M.\,H.~Stone \cite{Stone}, 1932}) There is a one-to-one correspondence between the linear self-adjoint operators $H$ in Hilbert space $\mathcal{F}$ and the unitary strongly continuous groups $(W(t))_{t\in \mathbb{R}}$ of linear bounded operators in $\mathcal{F}$. This correspondence is the following: $iH$ is the generator of $(W(t))_{t\in \mathbb{R}}$, which is denoted as $W(t)=e^{itH}.$ 
\end{theorem}

\begin{corollary}\label{cor} If $A$ is a linear self-adjoint operator in Hilbert space, then $\left\|e^{iA}\right\|=1.$ 
\end{corollary}

\begin{remark}\label{saop}Note that a linear self-adjoint operator in Hilbert space $\mathcal{F}$ by definition is closed and its domain is dense in $\mathcal{F}$. 
\end{remark}

The following Chernoff theorem allows to construct the $C_0$-semigroup in $\mathcal{F}$ from a suitable family of linear bounded operators in $\mathcal{F}$. This family usually does not have a semigroup composition property but is pretty close to a $C_0$-semigroup in the sense described in the theorem below. For many $C_0$-semigroups such families $G$ have been constructed, see \cite{Remizov1, BGS2010, Plya2, BB, Bmnog, Plya1, Butko1, Butko2, OSS}. Note that we present the Chernoff theorem in a new wording, see the motivation of that in chapter \ref{ha}.

\begin{theorem}\label{FormulaChernova}(\textsc{P.\,R.~Chernoff, 1968}; see \cite{Chernoff} or theorem 10.7.21 in \cite{BS}) Let $\mathcal{F}$ be a Banach space,
and $\mathcal{L}(\mathcal{F})$ be the space of all linear bounded operators in $\mathcal{F}$ endowed with the operator norm. Let $L\colon Dom(L)\to \mathcal{F}$ be a linear operator defined on $Dom(L)\subset\mathcal{F}$, and $G$ be an $\mathcal{L}(\mathcal{F})$-valued function. 

\textbf{Suppose} that $L$ and $G$ satisfy:

(E). There exists a $C_0$-semigroup $(e^{tL})_{t\geq 0}$ and its generator is $(L,Dom(L))$.

(CT1). The function $G$ is defined on $[0,+\infty)$, takes values in $\mathcal{L}(\mathcal{F})$, and the mapping $t\longmapsto G(t)f$ is continuous for every vector $f\in\mathcal{F}$. 

(CT2). $G(0)=I$. 

(CT3). There exists a dense subspace $\mathcal{D}\subset \mathcal{F}$ such that for every $f\in \mathcal{D}$ there exists
a limit $G'(0)f=\lim_{t\to 0}(G(t)f-f)/t$. 

(CT4). The operator  $(G'(0),\mathcal{D})$ has a closure $(L,Dom(L)).$ 

(N). There exists $\omega\in\mathbb{R}$ such that $\|G(t)\|\leq e^{\omega t}$ for all $t\geq 0$.

\textbf{Then} for every $f\in \mathcal{F}$ we have $(G(t/n))^nf\to e^{tL}f$ as $n\to \infty$, and the limit
is uniform with respect to $t\in [0,t_0]$ for every fixed $t_0>0$.
\end{theorem}

\begin{definition}\label{Cheq}Let $\mathcal{F}$ and $\mathcal{L}(\mathcal{F})$ be as before. Let us call two $\mathcal{L}(\mathcal{F})$-valued mappings $G_1$ and $G_2$ defined both on $[0,+\infty)$ (respectively, both on $\mathbb{R}$)  \textit{Chernoff-equivalent} if and only if  $G_1(0)=G_2(0)=I$ and for each $f\in\mathcal{F}$ and each~$T>0$
$$\lim_{n\to\infty}\sup_{\scriptsize\begin{array}{cc}
t\in [0,T]\\
(resp.\ t\in [-T,T]) \end{array}}\left\|\left(G_1\left(\frac{t}{n}\right)\right)^nf-\left(G_2\left(\frac{t}{n}\right)\right)^nf\right\|=0.$$
\end{definition} 

\begin{remark} There are several slightly different definitions of the Chernoff equivalence, we will just follow \cite{OSS} not going into details. The only thing we need from this definition is that if $G_1$ and $L$ satisfy all the conditions of the Chernoff theorem, then the mapping $G_1$ is Chernoff-equivalent to the mapping $G_2(t)=e^{tL}$. In other words, the limit of $(G_1(t/n))^n$ yields the $C_0$-semigroup $(e^{tL})_{t\geq 0}$ (or, respectively, $C_0$-group $(e^{tL})_{t\in\mathbb{R}}$) as $n$ tends to infinity.
\end{remark}

\begin{definition}\label{weakCher}Let us call a mapping $G$ \textit{Chernoff-tangent} to the operator $L$ iff it satisfies the conditions (CT1)-(CT4) of the Chernoff theorem.
\end{definition}

\begin{remark} With these definitions, the Chernoff-equivalence of $G$ to $(e^{tL})_{t\geq 0}$ follows from the existence (E) of the $C_0$-semigroup plus Chernoff-tangency (CT) plus the growth of the norm  bound (N). \end{remark}

\begin{remark}\label{corboundop} It is known that if $\mathcal{F}$ is a Banach space, and $A\colon \mathcal{F}\to \mathcal{F}$ is a linear bounded operator, then $e^A=\sum_{k=0}^\infty\frac{A^k}{k!}=\lim_{k\to\infty}\left(I+\frac{A}{k}\right)^k.$ Indeed, the operator $A$ is the generator of the $C_0$-semigroup $\left(e^{tA}\right)_{t\geq 0}$ defined by the formula $e^{tA}=\sum_{k=0}^\infty\frac{t^kA^k}{k!},$ see \cite{EN1} Chapter I, section 3: Uniformly continuous operator semigroups. Setting $L=A,$ $\mathcal{D}=\mathcal{F},$ $G(t)=I+tA$ and $\omega=\|A\|$ in theorem \ref{FormulaChernova} establishes the equality $e^{tA}=\lim_{n\to\infty}\left(I+\frac{tA}{n}\right)^n$ for all $t\geq 0$ and for $t=1$ in particular. 
\end{remark}

\begin{remark}\label{skorostskhod} The condition (CT3) of the Chernoff theorem says that $G(t)f=f + tLf +o(t)$ for each  $f\in \mathcal{D}$. It seems promising to try to find  $G(t)$ such that for fixed $k\in\mathbb{N}$ that $G(t)f=f + tLf +o(t^k)$, then one could expect a faster convergence $(G(t/n))^nf\to e^{tL}f$. 
\end{remark}

\section{Main theorem}\label{mt}

In what follows we consider the Schr\"odinger equation $i\psi'_t=\mathcal{H}\psi$ in the form 
$$\psi'_t=iaH\psi$$
where $a$ is a non-zero real constant and $H$ is a self-adjoint operator such that $aH=-\mathcal{H}$. Adding $a\neq 0$ to the formula helps to write the $C_0$-groups $(e^{itH})_{t\in\mathbb{R}}$ and $(e^{-itH})_{t\in\mathbb{R}}$ in one formula $(e^{iatH})_{t\in\mathbb{R}}$ just setting $a=1$ or $a=-1$, also $a$ may be used as a small or large parameter while $H$ is fixed. The motivation for this change of  notation is that the Chernoff theorem is usually applied to calculate  the operator $e^{tL}$ which is responsible for the solution to the equation $u'_t=Lu$.

\begin{theorem}\label{trick}\textbf{Suppose} that $\mathcal{F}$ is a complex Hilbert space having a dense linear subspace $Dom(H)\subset \mathcal{F}$. Suppose that a linear self-adjoint operator $H\colon Dom(H)\to \mathcal{F}$  and a non-zero number $a\in\mathbb{R}$ are given. Suppose that the mapping $S$ is Chernoff-tangent to $H$ and $(S(t))^*=S(t)$ for each $t\geq 0$. 

\textbf{Then} the family $\left(e^{ia(S(|t|)-I)\mathrm{sign}(t)}\right)_{t\in\mathbb{R}}$ is Chernoff-equivalent to the group $(e^{iatH})_{t\in\mathbb{R}}$ and for each fixed $t\in\mathbb{R}$ and $f\in\mathcal{F}$ the following holds: 

\begin{equation}\label{FFeyn1} e^{iatH}f=\lim_{n\to\infty}\left(e^{ia\big(S(|t/n|)-I\big)\mathrm{sign}(t)}\right)^nf=\lim_{n\to\infty}e^{ian\big(S(|t/n|)-I\big)\mathrm{sign}(t)}f,\end{equation}

\begin{equation}\label{FFeyn2} e^{iatH}f=\lim_{n\to\infty}\lim_{k\to\infty}\sum_{m=0}^k\frac{i^ma^mn^m(\mathrm{sign}(t))^m}{m!}\Big(S(|t/n|)-I\Big)^mf,\end{equation}

\begin{equation}\label{FFeyn4} e^{iatH}f=\lim_{n\to\infty}\lim_{k\to\infty}\left[\left(1-\frac{ian\, \mathrm{sign}(t)}{k}\right)I   + \frac{ian\, \mathrm{sign}(t)}{k} S(|t/n|)\right]^kf,\end{equation}

\begin{equation}\label{FFeyn3-newt} e^{iatH}f=\lim_{n\to\infty}\lim_{k\to\infty}\sum_{m=0}^k\sum_{q=0}^m\frac{(-1)^{m-q}i^ma^mn^m(\mathrm{sign}(t))^m}{q!(m-q)!} \Big(S(|t/n|)\Big)^q f,\end{equation}

\begin{equation}\label{FFeyn4-newt} e^{iatH}f=\lim_{n\to\infty}\lim_{k\to\infty} \sum_{q=0}^k  \frac{k!(k-ian\, \mathrm{sign}(t))^{k-q}(ian\, \mathrm{sign}(t))^q}{q!(k-q)!k^k} \Big(S(|t/n|)\Big)^q f, \end{equation}

\begin{equation}\label{FFeyn4-newtfull} e^{iatH}f=\lim_{n\to\infty}\lim_{k\to\infty} \sum_{m=0}^k \sum_{q=0}^{k-m} \frac{(-1)^{k-m-q}k!\, (ian\, \mathrm{sign}(t))^{k-q}}{m!q!(k-m-q)!k^{k-q}} \Big(S(|t/n|)\Big)^m f \end{equation}

where the limits are taken with respect to the norm in $\mathcal{F}.$  \end{theorem}

\textbf{Proof.} At first we obtain the above formulas for the case $t>0,$ i.e. $|t|=t$ and $\mathrm{sign(t)}=1$. Let us check the conditions of the Chernoff theorem for the $\mathcal{L}(\mathcal{F})$-valued mapping $R(t)=\exp(ia(S(t)-I))$ and the operator $iaH.$ 

For fixed $t>0$ the operator $ia(S(t)-I)$ is linear and bounded (recall (CT1) for $S$), so the exponent $e^{ia(S(t)-I)}$ is well-defined by the power series and the operator $e^{ia(S(t)-I)}$ is linear and bounded, see remark \ref{corboundop}. The continuity of $t\longmapsto R(t)$ in the strong operator topology follows from the continuity of $t\longmapsto S(t)$ in the strong operator topology and the continuity of the exponent in the norm topology. So (CT1) for $R$ is completed. (CT2) for $R$ follows from (CT2) for $S$: $R(0)=e^{ia(S(0)-I)}=e^{ia(I-I)}=e^0=I$. 

Let us prove (CT3) for $R$. Remember that (CT1) for $S$ says that for every $f\in \mathcal{F}$ the function $K_f\colon [0,+\infty)\ni t\longmapsto S(t)f\in\mathcal{F}$ is continuous. So by the Weierstrass extreme value theorem the set $K_f([0,1])\subset \mathcal{F}$ is compact and hence bounded for each $f\in\mathcal{F}$. This means that for each $f\in\mathcal{F}$ there exists a number $C_f>0$ such that $\|S(t)f\|\leq C_f$ for all $t\in[0,1]$. Next, by the Banach-Steinhaus uniform boundedness principle the family of linear bounded operators $(S(t))_{t\in[0,1]}$ is bounded collectively, i.e. there exists a number $C>0$ such that $\|S(t)\|<C$ for all $t\in[0,1]$. Suppose that linear operator $A\colon \mathcal{F}\to\mathcal{F}$ is bounded. Then $e^A=I+A+A^2\frac{1}{2!}+A^3\frac{1}{3!}+\dots=I+A+ A^2\sum_{n=0}^\infty\frac{A^n}{(n+2)!}\stackrel{denote}{=}I+A+ A^2\Psi(A).$ One can see that $$\|\Psi(A)\|=\left\|\sum_{n=0}^\infty\frac{A^n}{(n+2)!}\right\|\leq\sum_{n=0}^\infty\frac{\|A\|^n}{(n+2)!}\leq\sum_{n=0}^\infty\frac{\|A\|^n}{n!}=e^{\|A\|}.$$

Set $A=ia(S(t)-I)$. Then the estimates $\|A\|=\|ia(S(t)-I)\|\leq |a|(C+1)$ and $\Psi(ia(S(t)-I))\leq e^{|a|(C+1)}$ hold for all $t\in[0,1]$. So for all $t\in(0,1]$ we have
\begin{equation}\label{ct3forr}\frac{R(t)f-f}{t}=ia\frac{S(t)f-f}{t} -a^2 \Psi\Big(ia(S(t)-I)\Big)\, \Big(S(t)-I\Big)\, \frac{S(t)f-f}{t}.\end{equation}

Suppose that $f\in\mathcal{D}$ is fixed. Due to (CT3) for $S$ there exists a limit $\lim_{t\to 0}\frac{S(t)f-f}{t}=Hf,$ so $\frac{S(t)f-f}{t}=Hf+o(1).$  In the right-hand side of (\ref{ct3forr}) the last term for $t\in(0,1]$ can be estimated as follows: \
$$\left\|-a^2 \Psi(ia(S(t)-I))\ (S(t)-I)\ \frac{S(t)f-f}{t}\right\|\leq$$ 
$$|-a^2|\cdot \|\Psi(ia(S(t)-I))\|\cdot \left\|(S(t)-I)\frac{S(t)f-f}{t}\right\|\leq$$
$$|a|^2e^{|a|(C+1)}\|(S(t)-I)(Hf+o(1))\|\leq$$
$$|a|^2e^{|a|(C+1)}\Big(\|(S(t)-I)(Hf)\| +\|(S(t)-I)(o(1))\|\Big). $$

If $t\to 0$ then $\|(S(t)-I)(Hf)\|\to 0$ by (CT1) and (CT2) for $S$. Also $\|(S(t)-I)(o(1))\|\to 0$ because $\|o(1)\|\to 0$ and for $t\in(0,1]$ we have the norm bound $\|S(t)-I\|\leq C+1.$ So proceeding to the limit $t\to 0$ in (\ref{ct3forr}) we obtain $\lim_{t\to 0} \frac{R(t)f-f}{t}=ia\lim_{t\to 0}\frac{S(t)f-f}{t}=iaHf,$ which is (CT3) for $R$. 

[(CT4) for $S$]=[$(H,\mathcal{D})$ has the closure $(H,Dom(H))$]$\iff$ [$(iaH,\mathcal{D})$ has the closure $(iaH,Dom(H))$]=[(CT4) for $R$] because $Dom(H)=Dom(iaH)$.

By the Stone theorem the operator $(iaH,Dom(H))$ is the generator for the strongly continuous group $(e^{iatH})_{t\in\mathbb{R}}$ and of the strongly continuous semigroup $(e^{iatH})_{t\geq 0}$ in particular, so (E) for $R$ also holds. (N) with $\omega=0$ for $R$ follows from the condition $(S(t))^*=S(t)$ and the corollary \ref{cor}.

All the conditions of the Chernoff theorem for $R$ are fulfilled, which proves the first identity in (\ref{FFeyn1}). The second identity in (\ref{FFeyn1}) follows from the fact that $(e^A)^n=e^{nA}$ for each natural number $n$ and bounded operator $A$.  

To obtain (\ref{FFeyn2}) and (\ref{FFeyn4}) recall remark \ref{corboundop} which states for the bounded operator $A$ the equalities $e^A=\sum_{k=0}^\infty\frac{A^k}{k!}=\lim_{k\to\infty}(I+\frac{A}{k})^k$ and set $A=ian(S(t/n)-I)$ in (\ref{FFeyn1}). Applying the Newton binomial formula to (\ref{FFeyn2}) and (\ref{FFeyn4}), one obtains (\ref{FFeyn3-newt}) and (\ref{FFeyn4-newt}) respectively. Applying it to (\ref{FFeyn4-newt}) provides (\ref{FFeyn4-newtfull}).

Now let us go back to the general case $t\in\mathbb{R}$. Recall that $e^{iatH}$ exists for all real values of $t$ thanks to the Stone theorem. To prove (\ref{FFeyn1})-(\ref{FFeyn4-newtfull}) for $t<0$ substitute $t$ by $-t$, $a$ by $-a$ and apply the generation theorem for the groups from \cite{EN1} at p.79. The case $t=0$ is trivial. $\Box$

\begin{remark}Note that all the formulas stated in theorem \ref{trick} are not formal expressions. All the limits exist in $\mathcal{F}$, and this is an important part of the theorem's statement.
\end{remark}

\begin{remark}Note that in theorem \ref{trick} $f\in\mathcal{F}$ is fixed. The theorem does not state the uniform convergence of the limits with respect to $f\in\mathcal{F}$ or with respect to $f$ from some subset of $\mathcal{F}$. If $\mathcal{F}$ is a space of some functions $\mathcal{F}\ni f\colon Q\to\mathbb{C}$, $x\longmapsto f(x),$ then the theorem does not state the uniform convergence of the limits with respect to $x\in Q.$
\end{remark}

\begin{remark}If the operators $(S(t))_{t\geq 0}$ are integral operators, then the formulas obtained in the theorem above include both multiple integration (like Feynman formulas) and summation (not like Feynman formulas), this is why we propose to call them quasi-Feynman formulas. Such formulas give us one of the ways to solve the Cauchy problem for the equation $\psi'_t(t,x) =iaH\psi(t,x).$ \end{remark}

\begin{remark}\label{ststar} The conditions $S(t)=(S(t))^*$ and $H=H^*$ in the theorem above are not independent because the Chernoff tangency implies that $S(t)f=f+tHf+o(t)$ as $t\to 0$ for each $f$ from the core of $H$. 
\end{remark}

\begin{remark}\label{ststar1} If $S$ is Chernoff-tangent to $H$ but $S(t)\neq (S(t))^*$ for some $t$, one can substitute $S(t)$ by $(S(t)+(S(t))^*)/2$.
\end{remark}

\begin{remark}\label{remrost} One can put a polynomial of $S(t)$ into the exponent,
like $R(t)=\exp[i(a_0I+a_1S(t)+a_2(S(t))^2+\dots+a_n(S(t))^n)]$ or compute the values of $S(t)$ in several points like $R(t)=\exp[i(a_0I+a_1S(g_1(t))+\dots+a_nS(g_n(t)))]$ for the given functions $g_j\colon\mathbb{R}\to\mathbb{R}$ and numbers $a_j\in\mathbb{R}$, or combine these approaches.
\end{remark}

\begin{remark}\label{turkom}
 Yu.\,A.~Komlev and D.\,V.~Turaev have found the following application of the remarks \ref{remrost} and \ref{skorostskhod}. Let us consider $S(t)-I=\frac{S(t)-I}{t}t$ as a two-point finite difference approximation for $\frac{d}{dt}S(t)\big|_{t=0}$. Then, if we try e.g. a simple three-point approximation $\frac{d}{dt}S(t)\big|_{t=0}\approx \frac{1}{t}(-\frac{3}{2}I+2S(t)-\frac{1}{2}S(2t))$ then the family $R(t)=e^{ia\left(-\frac{3}{2}I+2S(t)-\frac{1}{2}S(2t)\right)}$ may give better Chernoff approximations to $e^{iatH},$ than $e^{ia(S(t)-I)}.$ One can also ask what will happen if we take a $d$-point approximation and then consider $d\to\infty.$
\end{remark}

\begin{remark}For a fixed $t$, the map $S(t)\colon f\longmapsto S(t)f$ is usually an integral operator over Gaussian measure. If one applies the finite difference approximation approach from remark \ref{turkom} directly to the function $f$, i.e. under the sign of the integral, the we can 
obtain a family $S(t)$ with $S(t)\big|_{t=0}=I, \frac{d}{dt}S(t)\big|_{t=0}=H, \frac{d^2}{dt^2}S(t)\big|_{t=0}=0,\dots, \frac{d^n}{dt^n}S(t)\big|_{t=0}=0$ by using fewer terms, because the Gaussian measure is symmetric.
\end{remark}

\begin{remark} Theorem \ref{trick} will be more useful if one proves that (at least in the most important cases) the limit in (\ref{FFeyn3-newt}), (\ref{FFeyn4-newt}), (\ref{FFeyn4-newtfull}) exists as a double limit, or at least that there exists a sequence $(k_n)$ of integers on which the limit $\lim\limits_{n\to\infty}\lim\limits_{k\to\infty}$ can be substituted by the limit $\lim\limits_{n\to\infty}$. 
\end{remark}

\section{Heuristic arguments}\label{ha}

It is usually not easy to construct a family which is Chernoff-equivalent to $(e^{itH})_{t\geq 0}$ because the conditions of the Chernoff theorem obstruct each other in some sense when dealing with a  Schr\"{o}dinger equation, especially in the case of infinite-dimensional configuration space $Q$. The main difficulties are: divergence of integrals -- one requires regularization, which is a change of the family $S$, i.e. a disturbing factor for the value at zero (CT2) and for the derivative at zero (CT3); proving the norm bound (N), which is associated with analytical difficulties and sometimes requires a change of the family $S$ or a change of the space $\mathcal{F}$; the selection of the space $\mathcal{F}$, which is connected to (CT1) and with all other conditions of the Chernoff theorem. An example of overcoming these difficulties one can find in \cite{Plya1} for the case $Q=\mathbb{R}^n$. Unfortunately in \cite{Plya1} the number $\varepsilon$ such that $0<\varepsilon<(2n+6)^{-1}$ is fixed and appears in a final Feynman formula, so the technique presented in \cite{Plya1} cannot be directly applied in the case of infinite-dimensional $Q$. 

However, in the case of the heat equation and the $C_0$-semigroup $(e^{tH})_{t\geq 0}$ the situation is usually simpler, e.g. Feynman formulas can be obtained for the case of infinite-dimensional $Q$ \cite{Remizov1, Remizov3}. So the initial idea (introduced in \cite{Remizov2}) was to use the family $(S(t))_{t\geq 0}$ which is Chernoff-equivalent to the $C_0$-semigroup $(e^{tH})_{t\geq 0}$ for constructing the family $(R(t))_{t\geq 0}$ which is Chernoff-equivalent to the $C_0$-semigroup $(e^{itH})_{t\geq 0}$. 

It helps to separate the conditions of the Chernoff theorem for $(R(t))_{t\geq 0}$ into independent blocks: existence of the $C_0$-semigroup (E) + Chernoff-tangency (CT) + growth of the norm bound (N). The first block is granted by the Stone theorem as $H$ is self-adjoint. The second block is achieved by arithmetic manipulations to save identity at zero and add $i$ to the derivative at zero. If we have an analytic function $r\colon\mathbb{C}\to\mathbb{C}$ with $r(0)=1$ and $r'(0)=i$ then we can define $R(t)=r(S(t))$. By choosing $r(z)=e^{i(z-1)}$ and $S(t)=(S(t))^*$ we can use the corollary \ref{cor} to obtain the third block. So we come to the formulas $R(t)=e^{i(S(t)-I)}$ and $e^{itH}=\lim_{n\to\infty}(R(t/n))^n$. 

After all we see that in the proof we do not need the Chernoff-equivalence of the family $(S(t))_{t\geq 0}$ to the $C_0$-semigroup $(e^{tH})_{t\geq 0}$, we need only the Chernoff-tangency of $(S(t))_{t\geq 0}$ to the operator $H$. Indeed, the proof holds on even if the $C_0$-semigroup $(e^{tH})_{t\geq 0}$ does not exist and the norm of $S(t)$ grows at any rate with respect to the growth of $t$. Thus, by allowing quasi-Feynman formulas instead of Feynman ones, a difficult task of a direct construction of the family which is Chernoff-equivalent to $(e^{itH})_{t\geq 0}$ is replaced by a simpler task of constructing a family which is Chernoff-tangent to $H$.  

Writing $e^{i(S(|t|)-I)\mathrm{sign}(t)}$ instead of $e^{i(S(t)-I)}$ arises as formal generalization step from the case $t\geq 0$ to the case $t\in\mathbb{R}.$

\section{Application scheme}\label{as}

As already mentioned, $C_0$-semigroups are used to study evolutionary equations
$u'_t(t,x)=Lu(t,x)$. Basic examples are heat equation $u'_t(t,x)=Hu(t,x)$ and the Schr\"{o}dinger equation $\psi'_t(t,x)=iH\psi(t,x)$. Here $t\in [0,+\infty)$ is time, and the spatial variable $x$ ranges over a set $Q$. 

Above we discussed a very general case as we worked only with $C_0$-semigroups and $C_0$-groups not taking into account what space $Q$ stays behind them. So the technique presented may, potentially, be employed in a case when $Q$ is $\mathbb{R}^n$ or some subset of $\mathbb{R}^n$, $\mathbb{C}^n$ or some subset of $\mathbb{C}^n$, a linear (Hilbert, Banach, etc.) space or some subset of it,  a lattice \cite{Ord}, a manifold of a finite or infinite dimension, a group, an algebra, a graph or ramified surface \cite{Dubravina}, etc. 

If one wants to do this, then $\mathcal{F}$ should be a complex Hilbert space of functions $f\colon Q\to\mathbb{C}$. With the method presented we can study equations for such functions $\psi\colon [0,+\infty)\times Q\to\mathbb{C}$ that for every fixed moment of time $t\in[0,+\infty)$ the function $x\longmapsto \psi(t,x)$ belongs to $\mathcal{F}$, and the function $t\longmapsto \psi(t,\cdot)$ is continuous and differentiable as a mapping $[0,+\infty)\to\mathcal{F}$. The discussion above does not lean on the nature of the scalar product in $\mathcal{F}$. For example, it can originate from the fact that $\mathcal{F}=L^2(Q,\mu)$ for some measure $\mu$ in $Q$, or it can be based on some other structures. As a very particular yet important case let us mention $Q=\mathbb{R}^3$ and $\mathcal{F}=L^2(\mathbb{R}^3)$ for the Schr\"{o}dinger equation.

As for the operator $H$, we need it to be linear and self-adjoint (hence densely defined and closed). For example, $H=\Delta$ or $H=\Delta^2$ or $(H\psi)(x)=(\Delta\psi)(x)-V(x)\psi(x)$ or some other. We need the coefficients of $H$ not to depend on $t$; nevertheless, they may depend on $x\in Q$.

Next, to construct a family $(S(t))_{t\geq 0}$ which is Chernoff-tangent to the operator $H$ in $\mathcal{F}=L^2(Q,\mu)$ one can use the following identities. They depend on $Q$ and we state them without details, just to sketch the idea. Denote a Gaussian measure \cite{DF, Go} in $Q$ with a correlation operator $B$ as $\mu_B$. Let $g\colon Q\to\mathbb{R}$ be a function bounded from zero and infinity plus some other properties, one can consider $g(x)\equiv \frac{1}{2}$ in this paragraph as a particular case. Let $V\colon Q\to\mathbb{R}$ be a function with $V(x)\leq 0$ and some other properties. Then the identities similar to $\int_Qf(x+y)\mu_{2tg(x)A}(dy)=f(x)+tg(x)\mathrm{trace}[Af''(x)]+o(t)$ and $e^{tV(x)}f(x)=f(x)+tV(x)+o(t)$ hold. If one denotes $(S(t)f)(x)=\int_Qf(x+y)\mu_{2tg(x)A}(dy)$, then $(S(t))_{t\geq 0}$ is Chernoff-tangent to $H=g(\cdot)\Delta$ as $(S(t)f)(x)=f(x)+tg(x)\Delta f(x)+o(t)$. If one denotes $(S(t)f)(x)=e^{tV(x)}\int_Qf(x+y)\mu_{2tg(x)A}(dy)$ then $(S(t))_{t\geq 0}$ is Chernoff-tangent to $H=g(\cdot)\Delta+V(\cdot)$ as $(S(t)f)(x)=f(x)+t[g(x)\Delta f(x) +V(x)f(x)]+o(t)$. See these and some other useful formulas (e.g. for $\nabla,$ Beltrami-Laplace operator, $(-\Delta)^n$) in more details with precise statements in \cite{Remizov1, BGS2010, Plya2, BB, Bmnog, Plya1, Butko1, Butko2, OSS, Dubravina}.

Now, suppose that all the above conditions are satisfied. Suppose that we have constructed a family $(S(t))_{t\geq 0}$ which is Chernoff-tangent to $H.$ Then the Cauchy problem in $\mathcal{F}$
$$
\left\{ \begin{array}{ll}
 \psi'_t(t,x)=iaH\psi(t,x);\quad t\in\mathbb{R}, x\in Q\\
 \psi(0,x)=\psi_0(x);\quad x\in Q\\
  \end{array} \right.
$$
stated for arbitrary $\psi_0\in \mathcal{F}$  and non-zero $a\in\mathbb{R}$ has the unique in $\mathcal{F}$ solution $\psi(t,x)=\left(e^{iatH}\psi_0\right)(x)$ depending on $\psi_0$ continuously with respect to the norm in $\mathcal{F}$, where for every $t\in\mathbb{R}$ the operator $e^{iatH}$ from the $C_0$-group $\left(e^{iatH}\right)_{t\in\mathbb{R}}$ in $\mathcal{F}$ is given by theorem \ref{trick}. If $\psi_0\in Dom(H)$, then the obtained solution is a strong solution, and in the general case $\psi_0\in\mathcal{F}$ it is a mild solution, see \cite{EN1} for definitions.

\section{Examples of application}\label{ex}
\subsection{One-dimensional equation with bounded smooth potential}

A.\,S.~Plyashechnik proposed a simple model to show how the method works and what sort of formulas for the solution it provides. The equation considered was previously solved by different methods (including Feynman formulas), but quasi-Feynman formulas are obtained for the first time.

 Suppose that a non-zero number $a\in\mathbb{R}$ and a differentiable function $V\in C_b^1(\mathbb{R},\mathbb{R})$ bounded along with its first derivative are given. Consider the Cauchy problem in $L^2(\mathbb{R}^1,\mathbb{C})$
 \begin{equation}
\left\{ \begin{array}{ll}\label{cauchyexample}
 \frac{i}{a}\psi_t'(t,x)=-\frac{1}{2}\psi_{xx}''(t,x) + V(x)\psi(t,x);\quad t\in\mathbb{R}, x\in \mathbb{R}\\
 \psi(0,x)=\psi_0(x);\quad x\in \mathbb{R}\\
  \end{array} \right.
\end{equation}
Let us rewrite it in the form
\begin{equation}
\left\{ \begin{array}{ll}
 \psi_t'(t,x)=iaH\psi(t,x);\quad t\in\mathbb{R}, x\in \mathbb{R}\\
 \psi(0,x)=\psi_0(x);\quad x\in \mathbb{R}\\
  \end{array} \right.
\end{equation}
where $H$ is an operator defined for $f\in W_2^2(\mathbb{R})$ by the formula $$(Hf)(x)=\frac{1}{2}f''(x) - V(x)f(x).$$ Here $W_2^2(\mathbb{R}) \subset L^2(\mathbb{R})$ is the Sobolev class, i.e. the linear space of all the functions $f\in L^2(\mathbb{R})$ such that $f'\in L^2(\mathbb{R})$ and $f''\in L^2(\mathbb{R})$ where $f'$ and $f''$ are the distributional derivatives of $f$. So in theorem \ref{trick} one can set $\mathcal{F}=L^2(\mathbb{R})$ and  $Dom(H)=W^2_2(\mathbb{R})$. This corresponds the case $Q=\mathbb{R}$ in section \ref{as}. 

The operator $S(t)$ is constructed as follows. Define $$(F_tf)(x)=\exp\left(-\frac{t}{2}V(x)\right)f(x)$$ and 
$$(B_tf)(x)=\frac{1}{\sqrt{2\pi t}} \int_{\mathbb{R}} e^{\frac{-(x-y)^2}{2t}}f(y)dy = \frac{1}{\sqrt{2\pi t}} \int_{\mathbb{R}} e^{\frac{-y^2}{2t}}f(x+y)dy$$ 
for $t>0$ and $B_0f=f.$ Then let us set $S(t)=F_t\circ B_t\circ F_t,$ i.e.
$$(S(t)f)(x)=\frac{1}{\sqrt{2\pi t}} \int_{\mathbb{R}} \exp\left(-\frac{y^2}{2t} -\frac{t}{2}\Big[V(x)+V(x+y)\Big]\right) f(x+y)dy.$$ 

It is not very difficult to check (this was done by D.\,V~Grishin and A.\,V.~Smirnov \cite{GS}) that all conditions of theorem \ref{trick} are fulfilled. To do this one can take the set $C_0^\infty(\mathbb{R},\mathbb{R})$ of all infinitely differentiable functions $\mathbb{R}\to\mathbb{R}$ with compact support for $\mathcal{D}$ in the definition of the Chernoff tangency, and then perform the calculations that are similar to what is done in \cite{Remizov3} in the proof of item 4 of theorem 4.1. 

Now take one of the formulas stated in theorem \ref{trick}, say, formula (\ref{FFeyn3-newt}): 
$$e^{iatH}f=\lim_{n\to\infty}\lim_{k\to\infty}\sum_{m=0}^k\sum_{q=0}^m\frac{(-1)^{m-q}(ian)^m(\mathrm{sign}(t))^m}{q!(m-q)!} \Big(S(|t/n|)\Big)^q f.$$
In our particular case it implies that the Cauchy problem (\ref{cauchyexample}) has defined for all $t\in\mathbb{R}$, the unique in $L^2(\mathbb{R})$ solution 

$$
\psi(t,x)=\lim_{n\to\infty}\lim_{k\to\infty}\sum_{m=0}^k\sum_{q=0}^m\frac{(-1)^{m-q}i^ma^mn^m(\mathrm{sign}(t))^m}{q!(m-q)!}\left(\frac{n}{2\pi |t|}\right)^{q/2}\times
$$
$$
\times \underbrace{\int\limits_\mathbb{R} \dots \int\limits_\mathbb{R}}_q \exp\left\{- \frac{|t|}{n}\left[\frac{1}{2}V(x)+\sum_{p=2}^qV\left(x+\sum_{j=p}^qy_j\right) + \frac{1}{2}V\left(x+\sum_{j=1}^qy_j\right) \right] \right\} \times
$$
$$
\times\exp\left[-\frac{n}{2|t|}\sum_{j=1}^q y_j^2\right]  \psi_0\left(x+\sum_{j=1}^q y_j\right)\prod_{p=1}^q dy_p.
$$

\subsection{Equation with a polyharmonic Hamiltonian}

Another example is provided by M.\,S.~Buzinov \cite{Buz}. Here we see the solution to the Cauchy problem for a type of Schr\"odinger equation that was not previously represented in a form of Feynman formula, but Feynman formulas were obtained for the corresponding heat equation \cite{BB}.

In this subsection we assume $Q=\mathbb{R}$. The function $V\colon \mathbb{R}\to\mathbb{R}$ is bounded and continuous. Arbitrary integer $N\geq 2$ is fixed. Consider the Cauchy problem for the higher order heat type parabolic equation
\begin{equation}\label{CP1}
 \left\{
   \begin{array}{ll}
      \frac{\partial}{\partial t}\omega(t,x) = -(-\Delta)^N\omega(t,x) - V(x)\omega(t,x); \quad t\in\mathbb{R}, x\in \mathbb{R},\\
     \omega(0,x)=\omega_0(x);\quad x\in \mathbb{R},
   \end{array}
 \right.
\end{equation}
and for the corresponding Schr\"odinger equation
\begin{equation}\label{CP2}
 \left\{
   \begin{array}{ll}
      i\frac{\partial}{\partial t}\psi(t,x) = (-\Delta)^N\psi(t,x) + V(x)\psi(t,x); \quad t\in\mathbb{R}, x\in \mathbb{R},\\
     \psi(0,x)=\psi_0(x);\quad x\in \mathbb{R}.
   \end{array}
 \right.
\end{equation}

Denoting $\mathcal{H}=(-\Delta)^N+V$ one can see that the equation (\ref{CP1}) can be rewritten as $\omega'_t=-\mathcal{H}\omega$. The equation (\ref{CP2}) can be rewritten as $i\psi'_t=\mathcal{H}\psi,$ which is the same as $\psi'_t=-i\mathcal{H}\psi.$ The Hamiltonian $\mathcal{H}$ is a self-adjoint operator $\mathcal{H}\colon Dom(\mathcal{H})\to L_2(\mathbb{R})$, where $Dom(\mathcal{H})$ is a domain of the closure in $L_2(\mathbb{R})$ of the operator $(-\Delta)^N$ initially defined on the Schwarz space $\mathcal{S}(\mathbb{R})\subset L_2(\mathbb{R})$.

\textbf{Feynman formulas for a heat-type equation}. Let us (following M.S.Buzinov) define for all $\omega\in L_2(\mathbb{R})$
$$(B(t)\omega)(x)=\left(l(t)\ast\omega\right)(x) =\int_{-\infty}^{+\infty}l(t,y)\omega(x-y)dy=\int_{-\infty}^{+\infty}l(t,y)\omega(x+y)dy$$
where for all $t>0$ and all $y\in \mathbb{R}$ we define $l(t,y)$ by the equality
$$(l(t))(y)=l(t,y)=l(t,-y)= \frac{1}{2\pi}\int_\mathbb{R}\frac{e^{ixy}dx}{1+tx^{2N}}$$
which can be rewritten as a finite sum using the Cauchy residue theorem:
$$\frac{1}{2\pi}\int_\mathbb{R}\frac{e^{ixy}dx}{1+tx^{2N}}= \left(Nt^\frac{1}{2N}\right)^{-1}\times $$
$$\sum\limits_{1\leq k\leq{\frac{N+1}{2}}}
\left[\alpha_k\cos\left(\beta_kt^\frac{-1}{2N}|y|\right)+\beta_k\sin\left(\beta_kt^\frac{-1}{2N}|y|\right)\right]
\exp\left(-\alpha_kt^\frac{-1}{2N}|y|\right)
$$
where for integer and positive $k$ is defined
$$
\alpha_k=\sin\frac{(2k-1)\pi}{2N},~~\beta_k=\cos\frac{(2k-1)\pi}{2N}.
$$

Let us also denote $(F(t)\omega)(x)=e^{-tV(x)}\omega(x).$ M.S.Buzinov proved \cite{Buz} that a) $B(t)^*=B(t)$ for all $t\geq 0$ b) $B(t)$ is Chernoff-equivalent to $e^{-t(-\Delta)^N}$ c)~$F(t)\circ B(t)$ is Chernoff-equivalent to $e^{-t\mathcal{H}}$ d) for all $\omega_0\in L_2(\mathbb{R})$ the Cauchy problem (\ref{CP1}) has the solution 
$$
\omega(t,x)=\left(e^{-t\mathcal{H}}\omega_0\right)(x) =\left(\lim_{n\to\infty}(F(t/n)\circ B(t/n))^n\omega_0\right)(x)=
$$ 

$$
\lim_{n\to\infty}\underbrace{\int_{-\infty}^{+\infty}\dots\int_{-\infty}^{+\infty}}_n \omega_0\left(x+\sum_{k=1}^ny_k\right)\left(\prod_{k=1}^nl\left(\frac{t}{n},y_k\right)\right)\times
$$

$$\exp\left[\frac{-t}{n}V(x)-\frac{t}{n}\sum_{k=1}^{n-1}V\left(x+\sum_{j=1}^ky_k\right) \right]dy_1\dots dy_n. 
$$

\textbf{Quasi-Feynman formulas for a Schr\"odinger equation.} Operators $F(t)$ and $B(t)$ are self-adjoint, but $F(t)\circ B(t)$ is not. Nevertheless we can define $(F_{1/2}(t)\omega)(x)=e^{-\frac{1}{2}tV(x)}\omega(x)$ and $S(t)=F_{1/2}(t)\circ B(t)\circ F_{1/2}(t)$ which provides $S(t)^*=S(t).$ This allows to employ theorem \ref{trick} and obtain the solution of (\ref{CP2}) in the form of a quasi-Feynman formula.

Indeed, let us set in theorem \ref{trick} $\mathcal{F}=L_2(\mathbb{R}),$ $S(t)=F_{1/2}(t)\circ B(t)\circ F_{1/2}(t),$ $H=\mathcal{H},$ $a=-1,$ $\mathcal{D}=\mathcal{S}(\mathbb{R})\subset L_2(\mathbb{R})$ -- the Schwarz space, and $Dom(\mathcal{H})$ as before in this subsection. We have
$$(S(t)\omega)(x)= e^{-\frac{1}{2}tV(x)}\int_{-\infty}^{+\infty}l(t,y)e^{-\frac{1}{2}tV(x+y)}\omega(x+y)dy.$$
Now we take one of the formulas from theorem \ref{trick}, say, formula (\ref{FFeyn3-newt}) and after calculation of the $(S(t/n)^q\omega)(x)$ arrive to the following solution for (\ref{CP2}):

$$\psi(t,x)=\lim_{n\to\infty}\lim_{k\to\infty}\sum_{m=0}^k\sum_{q=0}^m\frac{(-1)^{m-q}i^ma^mn^m(\mathrm{sign}(t))^m}{q!(m-q)!}$$

$$\underbrace{\int_{-\infty}^{+\infty}\dots\int_{-\infty}^{+\infty}}_q\psi_0\left(x+\sum_{j=1}^qy_j\right)\prod_{p=1}^ql\left(\frac{|t|}{n},y_p\right)\exp\left[\frac{-|t|}{2n}V(x)\right]$$

$$\exp\left[- \frac{|t|}{n}\sum_{p=1}^qV\left(x+\sum_{j=p}^qy_j\right) + \frac{|t|}{2n}V\left(x+\sum_{j=1}^qy_j\right)\right]dy_1\dots dy_q.$$

\section*{Acknowledgements}

Above all I am thankful to my first teacher of functional analysis, Professor Alexander V. Abrosimov \cite{Abrosimov}. I dedicate this paper to his blessed memory.

I thank my scientific advisor O.\,G.~Smolyanov for his active and kind criticism, and M.\,S.~Buzinov, A.\,V.~Duryagin, Yu.\,A.~Duryagina, D.\,A.~Fadeev, V.\,A~Filimonov, A.\,V.~Gorshkov, D.\,V.~Grishin, Yu.\,A.~Komlev, Yu.\,N.~Orlov, Ya.\,Yu.~Pavlovskii, A.\,S.~Plyashechnik, E.\,S.~Rozhkova, V.\,Zh.~Sakbaev, N.\,N. Sha\-marov, A.\,V.~Smirnov, D.\,V.~Turaev and guys from the MSU public forum forumlocal.ru  for helpful discussions and commenting on the manuscript. 

This work has been supported by the Russian Scientific Foundation Grant 14-41-00044.

\end{document}